\documentstyle[twoside]{article}

\catcode`\@=11
\long\def\@makefntext#1{
\protect\noindent \hbox to 3.2pt {\hskip-.9pt  
$^{{\eightrm\@thefnmark}}$\hfil}#1\hfill}		%CAN BE USED 

\def\@makefnmark{\hbox to 0pt{$^{\@thefnmark}$\hss}}	%ORIGINAL 
	
\def\ps@myheadings{\let\@mkboth\@gobbletwo
\def\@oddhead{\hbox{}
\rightmark\hfil\eightrm\thepage}   
\def\@oddfoot{}\def\@evenhead{\eightrm\thepage\hfil
\leftmark\hbox{}}\def\@evenfoot{}
\def\sectionmark##1{}\def\subsectionmark##1{}}

\oddsidemargin=\evensidemargin
\newcounter{sectionc}\newcounter{subsectionc}\newcounter{subsubsectionc}
\renewcommand{\section}[1] {\vspace{12pt}\addtocounter{sectionc}{1} 
\setcounter{subsectionc}{0}\setcounter{subsubsectionc}{0}\noindent 
	{\tenbf\thesectionc. #1}\par\vspace{5pt}}
\renewcommand{\subsection}[1] {\vspace{12pt}\addtocounter{subsectionc}{1} 
	\setcounter{subsubsectionc}{0}\noindent 
	{\bf\thesectionc.\thesubsectionc. {\kern1pt \bfit #1}}\par\vspace{5pt}}
\renewcommand{\subsubsection}[1] {\vspace{12pt}\addtocounter{subsubsectionc}{1}
	\noindent{\tenrm\thesectionc.\thesubsectionc.\thesubsubsectionc.
	{\kern1pt \tenit #1}}\par\vspace{5pt}}
\newcommand{\nonumsection}[1] {\vspace{12pt}\noindent{\tenbf #1}
	\par\vspace{5pt}}

\newcounter{appendixc}
\newcounter{subappendixc}[appendixc]
\newcounter{subsubappendixc}[subappendixc]
\renewcommand{\thesubappendixc}{\Alph{appendixc}.\arabic{subappendixc}}
\renewcommand{\thesubsubappendixc}
	{\Alph{appendixc}.\arabic{subappendixc}.\arabic{subsubappendixc}}

\renewcommand{\appendix}[1] {\vspace{12pt}
        \refstepcounter{appendixc}
        \setcounter{figure}{0}
        \setcounter{table}{0}
        \setcounter{lemma}{0}
        \setcounter{theorem}{0}
        \setcounter{corollary}{0}
        \setcounter{definition}{0}
        \setcounter{equation}{0}
        \renewcommand{\thefigure}{\Alph{appendixc}.\arabic{figure}}
        \renewcommand{\thetable}{\Alph{appendixc}.\arabic{table}}
        \renewcommand{\theappendixc}{\Alph{appendixc}}
        \renewcommand{\thelemma}{\Alph{appendixc}.\arabic{lemma}}
        \renewcommand{\thetheorem}{\Alph{appendixc}.\arabic{theorem}}
        \renewcommand{\thedefinition}{\Alph{appendixc}.\arabic{definition}}
        \renewcommand{\thecorollary}{\Alph{appendixc}.\arabic{corollary}}
        \renewcommand{\theequation}{\Alph{appendixc}.\arabic{equation}}
        \noindent{\tenbf Appendix \theappendixc #1}\par\vspace{5pt}}
\newcommand{\subappendix}[1] {\vspace{12pt}
        \refstepcounter{subappendixc}
        \noindent{\bf Appendix \thesubappendixc. {\kern1pt \bfit #1}}
	\par\vspace{5pt}}
\newcommand{\subsubappendix}[1] {\vspace{12pt}
        \refstepcounter{subsubappendixc}
        \noindent{\rm Appendix \thesubsubappendixc. {\kern1pt \tenit #1}}
	\par\vspace{5pt}}

\newcommand{\textlineskip}{\baselineskip=13pt}
\newcommand{\smalllineskip}{\baselineskip=10pt}

\def\eightcirc{
\begin{picture}(0,0)
\put(4.4,1.8){\circle{6.5}}
\end{picture}}
\def\eightcopyright{\eightcirc\kern2.7pt\hbox{\eightrm c}} 

\newcommand{\copyrightheading}[1]
	{\vspace*{-2.5cm}\smalllineskip{\flushleft
	{\footnotesize International Journal of Modern Physics A, #1}\\
	{\footnotesize $\eightcopyright$\, World Scientific Publishing
	 Company}\\
	 }}

\def\abstracts#1#2#3{{
	\centering{\begin{minipage}{4.5in}\baselineskip=10pt\footnotesize
	\parindent=0pt #1\par 
	\parindent=15pt #2\par
	\parindent=15pt #3
	\end{minipage}}\par}} 

\renewenvironment{thebibliography}[1]
	{\frenchspacing
	 \ninerm\baselineskip=11pt
	 \begin{list}{\arabic{enumi}.}
	{\usecounter{enumi}\setlength{\parsep}{0pt}
	 \setlength{\leftmargin 12.7pt}{\rightmargin 0pt} %FOR 1--9 ITEMS
	 \setlength{\itemsep}{0pt} \settowidth
	{\labelwidth}{#1.}\sloppy}}{\end{list}}

\newcounter{itemlistc}
\newcounter{romanlistc}
\newcounter{alphlistc}
\newcounter{arabiclistc}

\newcommand{\fcaption}[1]{
        \refstepcounter{figure}
        \setbox\@tempboxa = \hbox{\footnotesize Fig.~\thefigure. #1}
        \ifdim \wd\@tempboxa > 5in
           {\begin{center}
        \parbox{5in}{\footnotesize\smalllineskip Fig.~\thefigure. #1}
            \end{center}}
        \else
             {\begin{center}
             {\footnotesize Fig.~\thefigure. #1}
              \end{center}}
        \fi}

\newcommand{\tcaption}[1]{
        \refstepcounter{table}
        \setbox\@tempboxa = \hbox{\footnotesize Table~\thetable. #1}
        \ifdim \wd\@tempboxa > 5in
           {\begin{center}
        \parbox{5in}{\footnotesize\smalllineskip Table~\thetable. #1}
            \end{center}}
        \else
             {\begin{center}
             {\footnotesize Table~\thetable. #1}
              \end{center}}
        \fi}

\def\@citex[#1]#2{\if@filesw\immediate\write\@auxout
	{\string\citation{#2}}\fi
\def\@citea{}\@cite{\@for\@citeb:=#2\do
	{\@citea\def\@citea{,}\@ifundefined
	{b@\@citeb}{{\bf ?}\@warning
	{Citation `\@citeb' on page \thepage \space undefined}}
	{\csname b@\@citeb\endcsname}}}{#1}}

\newif\if@cghi
\def\cite{\@cghitrue\@ifnextchar [{\@tempswatrue
	\@citex}{\@tempswafalse\@citex[]}}
\def\citelow{\@cghifalse\@ifnextchar [{\@tempswatrue
	\@citex}{\@tempswafalse\@citex[]}}
\def\@cite#1#2{{$\null^{#1}$\if@tempswa\typeout
	{IJCGA warning: optional citation argument 
	ignored: `#2'} \fi}}

\def\pmb#1{\setbox0=\hbox{#1}
	\kern-.025em\copy0\kern-\wd0
	\kern.05em\copy0\kern-\wd0
	\kern-.025em\raise.0433em\box0}

\def\fnt#1#2{\footnotetext{\kern-.3em
	{$^{\mbox{\scriptsize #1}}$}{#2}}}

\def\fpage#1{\begingroup
\voffset=.3in
\thispagestyle{empty}\begin{table}[b]\centerline{\footnotesize #1}
	\end{table}\endgroup}

\def\runninghead#1#2{\pagestyle{myheadings}
\markboth{{\protect\footnotesize\it{\quad #1}}\hfill}
{\hfill{\protect\footnotesize\it{#2\quad}}}}
\font\tenrm=cmr10
\font\tenit=cmti10 
\font\tenbf=cmbx10
\font\bfit=cmbxti10 at 10pt
\font\ninerm=cmr9

\font\eightrm=cmr8

\def\qed{\hbox{${\vcenter{\vbox{			%HOLLOW SQUARE
   \hrule height 0.4pt\hbox{\vrule width 0.4pt height 6pt
   \kern5pt\vrule width 0.4pt}\hrule height 0.4pt}}}$}}

\newcommand{\beq}{\begin{equation}}
\newcommand{\eeq}{\end{equation}}

\begin{document}

\runninghead{Type IIB orientifolds with discrete torsion}
{Type IIB orientifolds with discrete torsion}
\normalsize\textlineskip
\thispagestyle{empty}
\setcounter{page}{1}

\copyrightheading{}			%{Vol. 0, No. 0 (1993) 000--000}

\vspace*{0.88truein}

\fpage{1}
\centerline{\bf TYPE IIB ORIENTIFOLDS WITH DISCRETE TORSION}
\vspace*{0.37truein}
\centerline{\footnotesize 
ROBERT L. KARP
\footnote{Present
address: Department of Mathematics, Duke University, Durham, NC 27708},
F. PAUL ESPOSITO, LOUIS WITTEN}
\centerline{\footnotesize\it Physics Department, University
of Cincinnati}
\baselineskip=10pt
\centerline{\footnotesize\it Cincinnati, Ohio 45221-0011,
USA}

\vspace*{0.225truein}

\vspace*{0.21truein}

\abstracts{We consider compact four-dimensional ${\bf Z_N}\times {\bf
Z_M}$ type IIB orientifolds, for certain values of $N$ and $M$. We allow
the additional feature of discrete torsion and discuss the
modification of the consistency conditions arising from tadpole
cancellation. We point out the differences between the cases with and
without discrete torsion. }{}{}

\textlineskip                   %) USE THIS MEASUREMENT WHEN THERE IS
\vspace*{12pt}                  %) NO SECTION HEADING
\noindent

Orientifold compactifications\cite{g} of the type IIB superstring
circumvent the problem that the type I theory does not produce a chiral
spectrum when compactified on a Calabi-Yau threefold with standard
embedding of the gauge degrees of freedom. Independently of this discrete
torsion (DT) was introduced as a phase factor related to the B-field,
allowed by modular invariance\cite{v} in orbifold compactifications of the
closed string theories. In the open string theories the analogous notion
of DT was discovered relatively recently, only after 
D-branes were better understood\cite{D}. In addition 
the relationship between closed and open DT 
has been further clarified\cite{o}.

The pioneering work for ${\bf Z_2}$
orientifolds was quickly generalized to ${\bf Z_n}$ for different values of
$n$'s\cite{gj}. The case ${\bf Z_2}\times {\bf Z_2}$ was
investigated\cite{bl} and generalized\cite{z}. The question of noncompact
orientifolds with DT was addressed as well\cite{kr}. The geometric aspects
of DT was partly described\cite{s} and there has recently been a revival of
interest in the subject\cite{n}.

The complete orientifold group we consider here is $G_1+ {\Omega} G_2$ with
${\Omega}h {\Omega} h' \in G_1$ for $h,h' \in G_2$. We restrict our
attention to $G_1=G_2={\bf Z_N}\times {\bf Z_M}$. The generator of either of
the factors will have the form $\theta= \exp (2i\pi
(v_1J_{45}+v_2J_{67}+v_3J_{89}))$, with $J_{mn}$ the $SO(6)$ Cartan
generators, acting on the compact $T^6$ (complexified) coordinates $Z_1 =
X_4+iX_5$, $Z_2 = X_6+iX_7$ and $Z_3 = X_8+iX_9$ as $\theta Z_i = {\rm
e}^{2i\pi v_i}Z_i$. If we chose the twist vectors of the ${\bf Z_N}$ and
${\bf Z_M}$ generators $\theta$ and $\omega$ to be of the form
$v_\theta=v= {\frac{1}N }(1,-1,0)$ and $v_\omega=w={ \frac{1}M}(0, 1,-1)$,
we end up with N=1 d=4 supersymmetry. Undoubtedly, there are many
equally interesting choices that do not have this form.

To derive the massless spectra we work in light-cone gauge. The GSO
projected untwisted massless Ramond states $| s_0 s_1 s_2 s_3 \rangle$
transform as $\theta |s_0 s_1 s_2 s_3 \rangle = {\rm e}^{2i\pi v\cdot s}
|s_0 s_1 s_2 s_3 \rangle$.

In this paper we will be mainly interested in the Klein bottle vacuum to
vacuum amplitude; the Mobius strip and the cylinder in fact have
similar expression. The Klein bottle amplitude is given by
\beq
{\cal K} = \frac{V_4}{2MN} \sum_{g,h}  
\int_0^\infty \frac{dt}{2t} \,
(4\pi^2 \alpha' t)^{-2} \,
{\rm Tr}_h \{{1+(-1)^F\over 2} \Omega \, g \,
{\rm e}^{-2\pi t[L_0(h) + \tilde{L}_0(h)]} \},
\eeq 
where the sums run over the entire group $G_1={\bf Z_N}\times {\bf Z_M}$,
and the trace is computed in the sector twisted by $h$. As any element of
$G_1$ is of the form $x^ay^b$, where $x$ ($y$) is a generator of ${\bf
Z_N}$ (${\bf Z_M}$), and $\Omega$ interchanges the sectors twisted by
$x^ay^b$ and $x^{N-a}y^{M-b}$, we see immediately that in order to have DT
make a difference in the Klein bottle amplitude we must require that either
$N$ or $M$ is even, and we have to study the sector twisted by $x^ay^b$,
with $(a,b)$ taken from the set $\{(N/2,0), (0,M/2),(N/2,M/2)\}$. Due to space
limitation we can only hint at the form of the expressions involved. For
example for the Klein bottle amplitude in the sector twisted by $x^{N/2}$
we have:
\beq
{\cal K}(x^{N\over 2},x^ay^b)=\chi_{({N\over 2},0)}^{(a,b)}\! 
\sum_{\alpha,\beta=0}^{1\over 2}\!
\eta_{\alpha,\beta}
\frac{\vartheta[{\alpha \atop \beta}]}{\eta^3} 
\left ( \prod_{i=1}^2
\frac{\vartheta[{{\alpha+{1\over 2}} \atop {\beta + 2u_i}}]}
{\vartheta[{0 \atop {{1\over 2} +2u_i}}]} \right )(-2\sin 2\pi u_3)
\frac{\vartheta[{\alpha \atop {\beta + 2u_3}}] }
{\vartheta[{{1\over 2} \atop {{1\over 2} +2u_3}}]},
\label{aa}
\eeq
where $u_i=av_i+bw_i$, and the argument of the $\vartheta$'s and $\eta$'s
is $e^{-4\pi t}$. In addition to this we also have factors coming from the
compact momenta and from windings\cite{g}. Using the properties of $\vartheta$
functions Eq. (\ref{aa}) can be simplified dramatically:
\beq
{\cal K}(x^{N\over 2},x^ay^b)=(1-1)\chi_{({N\over 2},0)}^{(a,b)}
\frac{\vartheta[{0 \atop {1\over 2}}]}{\eta^3}
\left ( \prod_{i=1}^2
\frac{\vartheta[{{{1\over 2}} \atop {{1\over 2}+ 2u_i}}]}
{\vartheta[{0 \atop {{1\over 2} +2u_i}}]} \right )(-2\sin 2\pi u_3)
\frac{\vartheta[{0 \atop {{1\over 2} + 2u_3}}] }
{\vartheta[{{1\over 2} \atop {{1\over 2} +2u_3}}]},
\eeq

In what follows we focus on the differences in the closed
string sector arising from the presence of DT.  For ${\bf Z_n}\times {\bf Z_m}$
with $n$ and $m$ odd, though DT is possible, it cannot contribute to the
Klein bottle amplitude, as the relevant twisted sector amplitudes are
zero. This implies that for these models there is no difference in the
tadpole cancellation equation with or without DT. As the orientifolds
constructed in these cases were based on projective representations on the
Chan-Paton indices, we conclude that DT has no effect.

The next simplest class of models is ${\bf Z_2}\times {\bf Z_m}$.  Here 
there are two sub-cases. For $m$ odd there is no DT. For $m$ even we can take
$x$ (resp. $y$) as the generator of ${\bf Z_2}$ (resp. ${\bf Z_m}$), and
$\omega_2=-1$ as the generator of $H^2({\bf Z_2}\times {\bf Z_m},U(1))$.
For the three potentially nonzero amplitudes we get: ${\cal K}(x,-)={\cal
K}(xy^{m/2},-)=0$, and ${\cal K}(y^{m/2},x^ay^b)\neq 0$ iff $2b/m\not\in
{\bf Z}$. Analyzing the different $m$'s is easy again. For $m=2$: $2b/m=b$
and this is integer, so all the twisted sector has zero amplitude. For
$m=4$ we have the $b=1,3$ nonzero amplitudes, but $\epsilon
(y^{2},x^ay^b)=1$, and DT has no effect. More generally for $m=4l$:
$\epsilon (y^{2l},x^ay^b)=1$ and we see that the ${\bf Z_2}\times {\bf
Z_{4l}}$ orbifold has the same tadpole cancellation condition with and
without DT. The case $m=6$ requires more work to see what happens.

For the ${\bf Z_2}\times {\bf Z_6}$ orientifold the sectors we are
interested in are the ones twisted by $x$, $xy^3$, $y^3$. It may easily
be seen that the sector twisted by $y^3$ is the only nonzero one. In what
follows we focus on the ${\cal K}(y^3,xy^a)=0$ contributions, which will
be proportional to $1/V_1$, as opposed to the ${\cal K}(y^3,y^a)$ contributions that are in
fact proportional to $V_1$. We also have $\epsilon (y^3,x^ay^b)=(-1)^a$.
It turns out that ${\cal K}(y^3,xy^a)=0$ for $a=0,3$, while for other
values of $a$ they all equal a common value proportional to
$\vartheta[{{1\over 2} \atop -{1\over 6}}] \vartheta[{{1\over 2} \atop
{1\over 6}}]/ \vartheta[{0 \atop -{1\over 6}}] \vartheta[{0 \atop {1\over
6}}]$. In the limit $t\rightarrow 0$ the twisted Klein bottle amplitudes
will give the contribution $(2t)\, (64\pi^2 \alpha')/V_1$ without DT, and
the negative of this with DT. Similarly, the untwisted Klein bottle
contribution that contributes with a factor of $1/V_1$ turns out to be
${\cal K}(1,xy^a)$, for $a=1,3,4,5$. In the $t\rightarrow 0$ limit these
add up the contribution $3\,(2t)\, (64\pi^2 \alpha')/V_1$. Thus in the
case without DT we have a tadpole contribution proportional to $(2t)\,
(256\pi^2 \alpha')/V_1$, which turns out to require $32$ D-branes to be
canceled. This agrees with the already known result\cite{z}. On the other
hand, for the case with DT the tadpole cannot be canceled, rendering the
model perturbatively inconsistent, in the sense of\cite{K}.

The next interesting case is ${\bf Z_3}\times {\bf Z_6}$. More generally
for $n$ odd the ${\bf Z_n}\times {\bf Z_{2n}}$ DT is $\epsilon
(y^{n},x^ay^b)=e^{(2\pi i/n)n(-b)}=1$, and once again DT has no effect.

The ${\bf Z_4}\times {\bf Z_4}$ model is interesting to analyze as well.
It is was known\cite{z} that without DT this model was perturbatively
inconsistent.  Our hope was that DT would change the tadpole cancellation
conditions, and allow for a consistent solution. It is elementary to show
that $\epsilon (x^2,x^ay^b)\neq 1$ iff $b=1,3$; $\epsilon
(x^2y^2,x^ay^b)\neq 1$ iff $a-b=-3,-1,1,3$, and $\epsilon (y^2,x^ay^b)\neq
1$ iff $a=1,3$. Unfortunately it turns out that with these constraints
${\cal K}(x^2,-)={\cal K}(x^2y^2,-)=0$, and ${\cal K}(y^2,-)=0$, implying
that even by turning on DT we cannot perturbatively save the model.

\vspace{12pt}

\noindent {\bf Acknowledgments}

\noindent
R.L.K. would like to thank R.G. Leigh for useful conversations. This work
was supported in part by the Department of Energy under the contract
number DOE-FGO2-84ER40153. RLK was also supported in part by the National
Science Foundation grant DMS-9983320.

\vspace{12pt}

\noindent {\bf Note}

\noindent
After this talk was given an exhaustive treatment of the subject
appeared\cite{Kl} that overlaps partly with our results.

\newpage
\nonumsection{References}

\end{document}